# Micro-cone arrays enhance outcoupling efficiency in horticulture luminescent solar concentrators

Zhijie Xu, Mark Portnoi, and Ioannis Papakonstantinou*

*Photonic Innovations Lab, Department of Electronic and Electrical Engineering, University College London, London WC1E 7JE, UK*
*Corresponding author: i.papakonstantinou@ucl.ac.uk*

**Luminescent solar concentrators (LSCs) have shown the ability to realize spectral conversion, which could tailor solar spectrum to better match photosynthesis requirements. However, conventional LSCs are designed to trap, rather than extract, spectrally converted light. Here, we propose an effective method for improving outcoupling efficiency, based on protruded and extruded micro-cone arrays patterned on the bottom surface of LSCs. Using Monte-Carlo ray tracing, we estimate a maximum external quantum efficiency (EQE) of 37.73% for our Horticulture-LSC (HLSC), corresponding to 53.78% improvement relatively to conventional, planar LSCs. Additionally, structured HLSCs provide diffuse light, beneficial for plant growth. Our micro-patterned surfaces provide a solution to light trapping in LSCs and a foundation for practical application of HLSC.**

http://dx.doi.org/10.1364/OL.99.099999

With the continuous increase of population, the demand for food is surging, putting more pressure on global agriculture [1]. Crop yields are significantly affected by photosynthesis efficiency, which is associated with photon spectrum [2]. In fact, only a narrow range of the solar spectrum benefits photosynthesis, mostly around the red wavelengths (600 nm - 700 nm) [3]. On the contrary, some bands, like the ultraviolet and green light, even induce unexpected damage to crops, leading to production decrease or quality reduction [4].

Unfortunately, the peak intensity distribution of AM1.5 does not match the preference for photosynthesis. The least efficient green component, occupies the peak and accounts for 35% of the photosynthetically active range (PAR, 400 nm - 700 nm) [5]. An idea to circumvent this disparity is by amplifying the red band at the expense of the green via luminescence. LSCs, consisting of polymer host matrices doped with fluorophores, can achieve this aim [6,7]. Fluorophores can absorb incident light and re-emit it by means of fluorescence [8,9]. By tuning the Stoke's shift, the re-emitted photon spectrum could be tailored to optimize photosynthesis [10–12].

Since the first invention of LSCs in the 1970s [13,14], the field has experienced rapid growth [15–17]. However, most LSCs devices are designed for photovoltaic applications [6,8], where it is beneficial to concentrate light by means of total internal reflection (TIR). Typically, light is emitted by the fluorophores isotropically. When the emission angle is greater than the critical angle, light falls within the TIR cone and is confined. For commonly used polymer hosts, a noticeable fraction (>70%) of photons end up being totally internally reflected [18]. In short, current LSCs are excellent photon concentrators, as their name suggests.

Consequently, even though LSC devices have shown potential for spectral conversion, if the outcoupling efficiency is poor, most of the converted photons would end up being trapped inside the polymer matrix and would never reach the plants. The Daily Light Integral (DLI), defined as the number of PAR photons per unit area received by plants within 24 hours, is a useful metric to evaluate crop yield [19]. Improving outcoupling efficiency, would increase the DLI, thereby promoting production in a greenhouse. Recent advances on light extraction techniques for optical displays and LEDs, provide possible avenues to boost outcoupling efficiency [20–23]. Intuitively, lowering the refractive index is the most obvious method to shrink the TIR cone and so, push more light out of the device. In recent studies, this was achieved by inducing subwavelength porosity within the polymer matrix, to create a medium with an effective index lower than its solid counterpart [24,25]. Alternatively, extrinsic structures, including microlenses [26], micro-pyramids [27] and micro-cones [28] were also reported to promote light extraction. Among them, micro-cone array achieves the best outcoupling efficiency on the condition of same area coverage and geometrical parameters [29].

In this letter, we draw inspiration from light extraction features in LEDs and optical displays and apply them in the context of LSCs to improve outcoupling efficiency. We used Monte-Carlo ray-tracing, a technique widely used in the LSC research [15,30–38], to analyse the performance of HLSCs (LightTools™, Optical Research Associates). In particular, we focus on hexagonally arranged micro-cone arrays on the bottom surface of LSCs, due to their ability to frustrate TIR and deflect photons out of lightguides. Both extruding and protruding features are modelled and discussed. Effects of fluorophore concentration, HLSC thickness and cone height/radius ratio (H/R) on photon fates are investigated.

Before discussing the modelling results, some metrics should be established for evaluating outcoupling efficiency. Note that the metrics used here are not the same as the ones in conventional LSCs, reflecting the different nature of our problem. Firstly, internal quantum efficiency (IQE) is defined as the ratio of photons escaping from the *bottom* surface to the total number of photons absorbed, see Figure 1. This parameter characterizes the ability to extract the converted photons from the bottom surface, which is the useful surface emitting towards the plants (photons leaving from the top and side surfaces are considered lost). The second parameter is the external quantum efficiency (EQE), which in our context, is defined as the ratio of photons escaping from the bottom surface to the *total* number of incident photons. EQE marks the total utilization rate of solar energy.

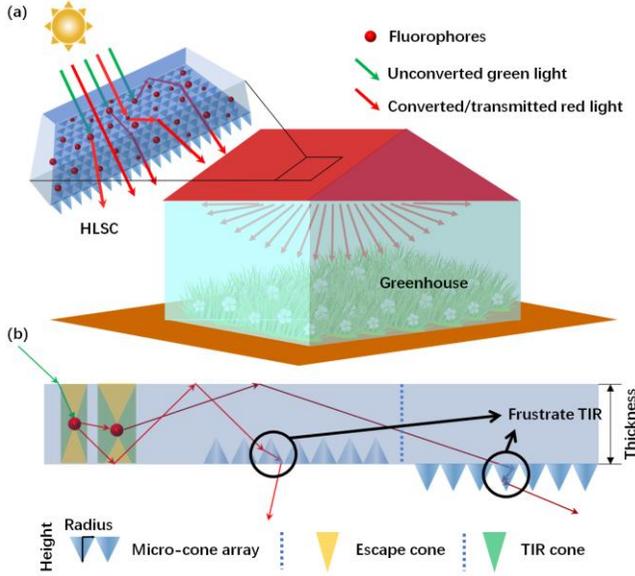

Fig. 1. (a) Schematic representation of a greenhouse HLSC with light-extraction structures. (b) Concept of HLSC improving light extraction. Green light is absorbed by fluorophores, then converted into red light. Re-emitted light that fails to fall into escape cone (orange triangle) would be trapped in a planar LSC. Protruding and extruding micro-cone arrays (blue triangles) frustrate TIR, however, and improve outcoupling efficiency.

Figure 1(a) illustrates the schematic diagram of HLSC, whereby green light is first red-shifted before being emitted towards the interior of the greenhouse to facilitate photosynthesis. In our simulations, we used an artificial polymer matrix with a refractive index of 1.5 (common for polymers used in LSC research, [18]) and a prototypical fluorescent dye; Lumogen Red, an economical dye with high quantum yield (96%), widely adopted in LSC research. The absorption spectrum of Lumogen Red peaks at 575 nm, and the re-emitted light spans the wavelength range from 570 nm to 700 nm, [39], making it ideal for photosynthesis enhancement.

Figure 1(b) illustrates the mechanism of light extraction. Re-emitted photons reach the bottom, polymer-air interface. If the incident angle is smaller than the critical angle $\theta_c$, defined by Snell's law: $\theta_c = \sin^{-1}(\frac{n_{air}}{n_{LSC}})$ where $n_{air}$ and $n_{LSC}$ are refractive indices of air and host matrix correspondingly, the photons fall into the escape cone (depicted by orange triangles in Figure 1b) and leave the device. Otherwise, photons are trapped due to TIR (green triangle in Figure 1(b)). Considering the refractive index is 1.5 in our simulations, the corresponding critical angle is ~42°. Accounting for the solid angle subtended by the TIR cone, the portion of trapped photons is given by: $\eta = \frac{\int_0^{2\pi}\int_{\theta_c}^{\pi-\theta_c}\sin(\theta)d\theta d\phi}{4\pi} = \cos(\theta_c)$. As a result, the trapping efficiency is ~74%. To reduce light trapping, micro-extraction features patterned on the bottom surface are proposed to frustrate the TIR process in the HLSC device. As shown in Figure 1(b), both protruding and extruding micro-cone arrays are considered, as both can easily be made by scalable fabrication methods (such as nanoimprint lithography, roll-to-roll hot-embossing or other [40]). The key geometrical and material parameters to examine are the H/R, the thickness of the device and the fluorophore concentration in the host polymer. The performance of sparser, non-touching base cone arrays is consistently worse than this with touching bases, due to the lower coverage area. So, to achieve highest density, all micro-cones are arranged in a hexagonal manner with touching bases.

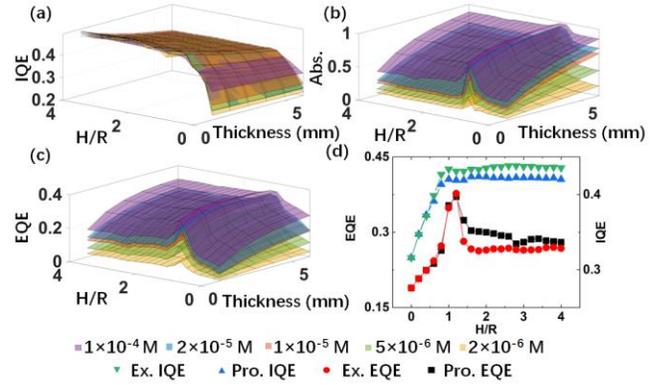

Fig. 2. Simulated (a) IQE, (b) absorbance (Abs.) and (c) EQE as a function of H/R, HLSC thickness and fluorophore concentration for extruding structures. (d) Comparison of IQE and EQE for extruding and protruding structures.

Based on the definitions of EQE and IQE, the following relationship can be derived connecting these two key metrics: $\frac{EQE}{IQE} = (1 - \frac{Fresnel\ Loss}{Input\ photons}) \times Absorbance$. Fresnel loss accounts for ~4%, when the refractive index is 1.5 and the incident angle is normal [41], so the first term on the right hand side of the equation is ~1. EQE, thus, is approximately equal to the product of the absorbance and IQE.

To optimize EQE, we chose to work with a monochromatic light source at 520 nm. While this wavelength falls in the absorption spectrum, it avoids the emission spectrum, which is convenient for photon fate calculation, as we have shown in our previous work [15]. The area of the HLSC was fixed to 80×80 mm (thickness varying from 1 to 6 mm), which is compatible with small scale experimental prototypes [34], and, also, because results remain essentially the same for larger dimensions. The fluorophore concentration used in our simulations ranged from 2×10$^{-6}$ M to 1×10$^{-4}$ M, equivalent to absorbance ranging from 0.026 to 0.855 through the LSC thickness. Besides, our simulations verified that EQE remains invariant to radius changes from 10 μm to 1 mm, when the H/R ratio is maintained constant. As such we decided to fix the radius to 50 μm, a value easily attainable.

As shown in Figure 2(a), IQE shows a monotonic increase with H/R in the examined range from 0 to 4. Note that H/R=0, implies a planar device and is used here for comparison with the micro-cones. H/R=4 corresponds to a very sharp cone and was taken as the limit, since larger values didn't show further improvement. A plateau actually starts appearing already at H/R~0.8, indicating that light extraction efficiency is saturated. This is the critical value, beyond which, most TIR rays meet the micro-cone facets at a sufficiently acute angle to cause their extraction, see Figure 1(b). As all converted photons are assumed to be emitted isotropically, IQE depends more on the geometrical parameters of the micro-cones and less on the device thickness and fluorophore concentration. The highest IQEs for 1, 2, 3, 4, 5, 6 mm thick samples are 49%, 45%, 43%, 41%, 39% and 38%, correspondingly. IQE with a maximum increase of about 119%, compared with a planar LSC, is achieved for H/R=3 when the concentration is $2\times10^{-6}$ M and thickness is 1 mm.

Variation of absorbance shows a distinctly different trend. As shown in Figure 2(b), thickness, fluorophore concentration and H/R all noticeably affect absorbance. Increasing fluorophore concentration, for example, reduces the mean free path for absorption. As a result, the probability that a photon is absorbed increases. Thicker samples, on the other hand, lengthen the optical path light travels in the device, amplifying, again, photon absorbance. Besides, the cone structure may also back-reflect some unabsorbed photons, providing them with yet another chance to be absorbed. Take the example of near normal incidence. When H/R = 1 (cone tip angle of 90°), most unabsorbed photons reaching the bottom textured surface will be back-reflected, doubling the optical path in the device. In this case, significant improvements compared to the planar LSC can be attained, as exemplified in Figure 2(b).

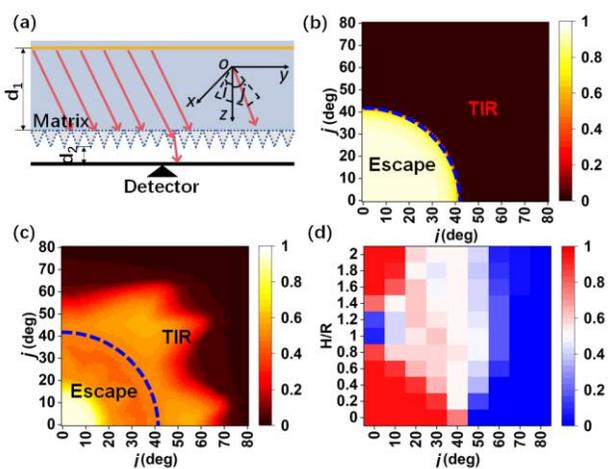

Fig. 3. (a) Simulation model used for testing micro-cone arrays enhancing light extraction. Outcoupling efficiency as a function of incident angle detected from (b) planar polymer slab and (c) micro-cone array slab for H/R=1.2. The blue dashed line delineates the end of the escape cone and the beginning of the TIR cone. Only a quadrant is simulated due to the symmetry of cones. (d) Detected outcoupling efficiency for different incident angles $i$ and H/R.

Benefiting from the increases in IQE and absorbance, EQE also improves significantly by the micro-cone arrays. According to Figure 2(c), EQE closely tracks the variation of absorbance, particularly for the higher H/R for which IQE saturates. The highest EQE achieved is 37.73% for H/R=1.2, concentration $1\times10^{-4}$ M and sample thickness 3 mm. This corresponds to a 53.78% improvement compared with the best results obtained with a planar LSC. According to simulation results from Figure 2(d), the protruding structures show a similar performance with the extruding, offering flexibility in future HLSC device fabrication.

To shed more light into how micro-cones frustrate TIR, we run the additional simulations summarized in Figure 3. In this case, we positioned a plane wave source inside a micro-cone textured polymer slab which was clear (i.e., not doped with fluorophores). Then, we monitored the portion of photons escaping the bottom surface as a function of the incident angle, by placing a detector just underneath the micro-cones, Figure 3(a). Note that H/R is set as 1.2 here, the value at which highest EQE is achieved. At the top of the polymer slab, a perfectly absorbing layer was placed to stop all rays not extracted by the micro-cones and remove them from the simulation region. For comparison, we repeated the same simulations for a planar slab. Outcoupling efficiency (i.e., ratio between photons escaping from a certain incident angle over initial number of photons emitted at this angle) is plotted in Figure 3(b) and Figure 3(c) as a function of incident angles ($i,j$). $i$ and $j$ represent the orientations in X-Z and Y-Z plane respectively (as shown in Figure 3a). According to Figure 3(b), only incident angles that are smaller than the critical angle escape in the planar slab case (escape zone), as obviously expected. The behavior of micro-cones is strikingly different though. According to Figure 3(c), the angular distribution of rays exiting the textured surface extends far into the TIR zone (TIR limits are represented by the blue-dashed line). As a matter of fact, the outcoupling efficiency is sustained to >0.4 for incident angles even greater than 60°, and it never drops to zero even for nearly grazing angles. This expansion of extracted angles beyond the escape cone is the principal mechanism of outcoupling efficiency improvement in our system. Variation of outcoupling efficiency based on H/R and incident angle $i$ ($j$ is fixed as 0°) is concluded in Figure 3(d). For all H/R, extraction angles always extend beyond TIR, verifying that micro-cones is an effective way to improve outcoupling efficiency.

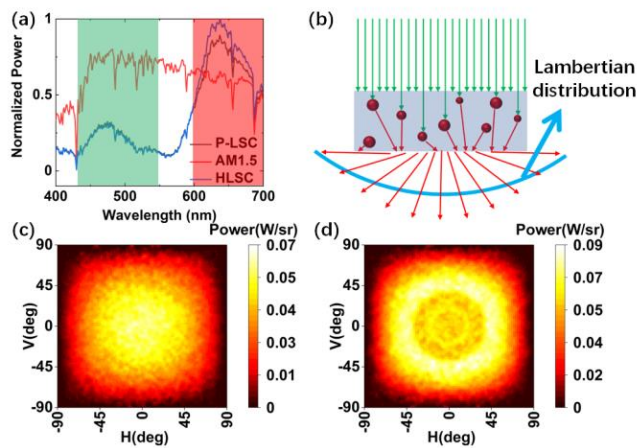

Fig. 4. (a) Normalized radiant power escaping planar-LSC and micro-cone HLSC and comparison with AM1.5 spectrum. Green area represents spectral band of poor photosynthesis efficiency, while red area represents high. (b) Concept of HLSC providing diffuse light for plant growth. Simulated angular distribution of radiant intensity for (c) planar-LSC and (d) micro-cone HLSC. Angles V and H are based on Type B coordinate system for Goniophotometer.

Once micro-cones were optimized for optimum outcoupling efficiency, new simulations were ran, to account for the entire AM1.5 spectrum. Figure 4(a) shows the spectral distribution of light escaping from the bottom of the device for i) a planar-LSC and ii) a micro-cone HLSC. All results were normalized to the peak value of the micro-cone HLSC spectrum. In these simulations, the thickness of the device was 5 mm and fluorophore concentration was $1\times10^{-4}$ M, corresponding to the values showing best performance. Both planar-LSC and micro-cone HLSC show a significant reduction of the power in the green band and concomitant energy transfer to the red. Compared with its planar counterpart, a micro-cone HLSC shows an increased power of 11.8% in the red. However, this improvement is somewhat lower than that of EQE (~53.78%, as discussed before). This is attributed to the back-reflection from unabsorbed red component in AM1.5 spectrum. Despite this, power in red area from structured HLSC is still 52.58% higher than that from AM1.5 spectrum.

Finally, it is noted that diffuse light is preferable, as it can reach not only the leaves but also the stems and roots of plants, [42], Figure 4(b). And as found, micro-cone HLSC acts as excellent light diffuser too. Figure 4(c) shows the angular distribution of light escaping from a planar-LSC. Figure 4(d), on the other hand, illustrates the angular profile of photons escaping the micro-cone HLSC. For both cases, the angular profile resembles a Lambertian distribution, showing that the extra outcoupling efficiency in micro-cone HLSCs, is not impacting photon randomization.

In summary, micro-cone arrays on the bottom surface of LSCs were optimized for light extraction using Monte-Carlo ray tracing method. The morphology of the micro-cone arrays was investigated by controlling the H/R to obtain best IQE and absorbance, thereby facilitating the improvement of EQE. Combined with sample thickness and fluorophore concentration optimization, the highest EQE of 37.73% could be achieved which corresponds to 11.8% more red light reaching the plants compared to an equivalent planar LSC. Simultaneous spectral conversion, enhanced light extraction and light diffusion were realized with our micro-cone HSLC device, showing great potential for horticulture applications. In the future, efforts will be focused on designs that do not affect the original red spectrum, for example by texturing both surfaces of the LSC or using asymmetric cone geometries.

**Funding.** We are grateful to the Chinese Scholarship Council (CSC) for the award of a PhD studentship and UCL Faculty of Engineering for the award of a Dean's prize.

**Disclosures**. The authors declare no conflicts of interest.